\documentclass[conference]{IEEEtran}
\IEEEoverridecommandlockouts
\usepackage{cite}
\usepackage{amsmath,amssymb,amsfonts}
\usepackage{graphicx}
\usepackage{textcomp}
\usepackage{xcolor}
\usepackage{booktabs}
\usepackage{algorithm}
\usepackage{algorithmic}
\usepackage{multirow}
\usepackage{enumitem}
\usepackage{pifont}
\usepackage{hyperref}
\hypersetup{
    colorlinks=true,
    linkcolor=blue,
    filecolor=magenta,
    urlcolor=cyan,
    citecolor=cyan
}

\def\BibTeX{{\rm B\kern-.05em{\sc i\kern-.025em b}\kern-.08em
    T\kern-.1667em\lower.7ex\hbox{E}\kern-.125emX}}
\begin{document}

\title{BiasIG: Benchmarking Multi-dimensional\\Social Biases in Text-to-Image Models}

\author{
\IEEEauthorblockN{
Hanjun Luo\textsuperscript{1,\ddag},
Zhimu Huang\textsuperscript{1,*},
Haoyu Huang\textsuperscript{2,*},
Ziye Deng\textsuperscript{2,*},
Ruizhe Chen\textsuperscript{2},
\\
Xinfeng Li\textsuperscript{3},
Zuozhu Liu\textsuperscript{2,\dag},
Hanan Salam\textsuperscript{1}
}
\IEEEauthorblockA{\textsuperscript{1}New York University Abu Dhabi \quad
\textsuperscript{2}Zhejiang University \quad
\textsuperscript{3}Nanyang Technological University}
\IEEEauthorblockA{\textsuperscript{*}Equal contribution \quad \textsuperscript{\dag}Corresponding author \quad \textsuperscript{\ddag}hl6266@nyu.edu}
}
\maketitle

\begin{abstract}
Text-to-Image (T2I) generative models have revolutionized content creation, yet they inherently risk amplifying societal biases. While sociological research provides systematic classifications of bias, existing T2I benchmarks largely conflate these nuances or focus narrowly on occupational stereotypes, leaving the \emph{multi-dimensional nature} of generative bias inadequately measured. In this paper, we introduce \textbf{\texttt{BiasIG}}, a unified benchmark that quantifies social biases across a curated dataset of \textit{47,040} prompts. Grounded in sociological and machine ethics frameworks, \texttt{BiasIG} disentangles biases across \textit{4} dimensions to enable fine-grained diagnosis. To facilitate scalable and reliable evaluation, we propose a fully automated pipeline powered by a fine-tuned multi-modal large language model, achieving high alignment accuracy comparable to human experts. Extensive experiments on \textit{8} T2I models and \textit{3} debiasing methods not only validate \texttt{BiasIG} as a robust diagnostic tool, but also reveal critical insights: interventions on protected attributes often trigger unintended confounding effects on unrelated demographics, and debiasing methods exhibit a persistent tendency toward \emph{discrimination} rather than mere \emph{ignorance}. Our work advocates for a precise, taxonomy-driven approach to fairness in AIGC, providing a theoretical framework for using \texttt{BiasIG}'s metrics as feedback signals in future closed-loop mitigation. The benchmark is openly available at \url{https://github.com/Astarojth/BiasIG}.
\end{abstract}

\begin{IEEEkeywords}
text-to-image, social bias, benchmark, fairness
\end{IEEEkeywords}

\section{Introduction}

Text-to-image (T2I) models are reshaping visual creation. However, their reliability is undermined by \emph{Social Bias}, the systematic deviations from user intent and sociological reality \cite{wan2024survey,sufian2025t2ibias,luo2024faintbench}. In practice, this often appears as (i) \emph{implicit} defaults to stereotypical representations under underspecified prompts (e.g., rendering ``CEO'' exclusively as white males) \cite{bianchi2023easily}, and (ii) \emph{explicit} failures to follow protected-attribute instructions (e.g., ignoring ``an Asian husband and white wife''). These failures not only reflect distributional mismatch, but also reinforce harmful stereotypes, erase plausible demographic presence, and undermine representational dignity in AIGC systems.

\vspace{-0.7em}
\begin{table}[H]
  \centering
    \caption{Summary and comparison of existing benchmarks.}
    \vspace{-0.2em}
  \label{tab:bench_compara}
  \resizebox{0.43\textwidth}{!}{
  \begin{tabular}{lccccc}
        \toprule
        \textbf{Benchmark} & Model & Prompt & Metric & Multi-level \\
        \midrule
        \textbf{DALL-Eval \cite{cho2023dall}} & 4 &  252 & 6 & no\\
        \textbf{HRS-Bench \cite{bakr2023hrs}} & 5 & 3000 & 3 & no \\
        \textbf{ENTIGEN \cite{bansal2022well}} & 3 & 246 & 4 & yes \\
        \textbf{TIBET \cite{chinchure2023tibet}} & 2 & 100 & 7 & no \\
        \textbf{T2ISafety \cite{li2025t2isafety}} & 15 & 236 & 1 & yes \\
        \textbf{MAGBIG \cite{friedrich2025multilingual}} & 5 & 3630 & 3 & no \\
        \midrule
        \textbf{\textbf{\texttt{BiasIG}}} & 8+3 & 47040 & 18 & yes \\
        \bottomrule
    \end{tabular}
    }
\end{table}
\vspace{-0.7em}

Despite emerging mitigation efforts \cite{luo2024versusdebias,cai2025autodebias}, existing T2I benchmarks remain fragmented in three ways: (i) \textit{Restricted Coverage}, with prompt sets largely centered on occupational stereotypes; (ii) \textit{Fragmented Evaluation}, where implicit diversity and explicit instruction failure are measured separately; and (iii) \textit{Lack of T2I-Specific Taxonomy}, where general ML bias notions are imported without capturing the distinction between generative ignorance and discrimination.

To bridge these gaps, we introduce \textbf{\texttt{BiasIG}}, a unified benchmark for systematically quantifying social \textbf{Bias}es in \textbf{I}mage \textbf{G}eneration. Table~\ref{tab:bench_compara} compares \textbf{\texttt{BiasIG}} with existing benchmarks. Grounded in sociological and machine-ethical frameworks, we define a four-dimensional taxonomy: \textit{Acquired Attributes}, \textit{Protected Attributes}, \textit{Manifestation}, and \textit{Visibility}. This structured approach allows us to synthesize 47,040 prompts, spanning occupations, personal characteristics, and complex social relations. For scalable and precise auditing, we implement a fully automated, human-validated evaluation pipeline with a fine-tuned Multi-Modal Large Language Model (MLLM), leveraging visual reasoning to accurately measure both implicit and explicit biases. We apply \textbf{\texttt{BiasIG}} to benchmark \textbf{8} T2I models and \textbf{3} debiasing methods. Our empirical analysis moves beyond simple ranking, uncovering structural phenomena, such that interventions on one demographic skew others and a persistent tendency toward discrimination. Our contributions are summarized as follows: 

\begin{itemize}[
    leftmargin=*,     
    topsep=-0.5em,       
    partopsep=0pt,    
    parsep=0pt,      
    itemsep=0pt       
]
    \item[\ding{182}] \textbf{\textit{4D Definition.}} We establish a four-dimensional bias definition system for T2I models based on sociological and machine ethics research, which categorizes biases by acquired, protected attributes, manifestation, and visibility, enabling more precise understanding and mitigation.
    
    \item[\ding{183}] \textbf{\textit{Unified Benchmark.}} We present \textbf{\texttt{BiasIG}}, a unified benchmark for evaluating T2I model biases. It features a 47,040-prompt dataset and an automated, high-accuracy evaluation pipeline, providing a versatile and efficient research tool.

    \item[\ding{184}] \textbf{\textit{Empirical Findings.}} We evaluate 8 T2I models and 3 debiasing methods, revealing critical structural limitations and outlining a theoretical roadmap based on our findings. 
\end{itemize}

\section{Definition System}
\label{define}

To dismantle bias conflation in existing benchmarks, we propose a tailored taxonomy grounded in sociological and machine-ethical frameworks \cite{landy2018bias,varona2022discrimination}. We operationalize social bias in T2I models into a four-dimensional structure: \textit{acquired attributes}, \textit{protected attributes}, \textit{manifestation}, and \textit{visibility}. This system maps any generative bias to a coordinate within this four-dimensional space.

\paragraph{Acquired Attributes (Context)}
These represent mutable traits derived from individual experience, socioeconomic status, or choices (e.g., \textit{occupation}, \textit{social relation}). While serving as legitimate bases for differentiation in real-world contexts, in generative models, they function as the semantic locus where stereotypical associations are frequently triggered.

\paragraph{Protected Attributes (Identity)}
These denote immutable or legally protected group identities (e.g., \textit{race}, \textit{sex}) that serve as the demographic variables for fairness auditing. Ethically, these attributes should remain statistically decoupled from acquired attributes to ensure representational equity.

\paragraph{Manifestation of Bias}
Drawing on social psychology \cite{devine1989stereotypes}, we operationalize social bias manifestation into two modes based on output distribution, bridging the gap between sociological concepts and machine learning mechanics:
\begin{itemize}[leftmargin=*,
    topsep=0em,
    partopsep=0pt,
    parsep=0pt,
    itemsep=0pt
    ]

\item[\ding{224}] \textit{Ignorance}
represents a state of \emph{representational homogeneity}, where models consistently generate a dominant demographic group regardless of semantic context. From a statistical learning perspective, ignorance arises when specific groups are severely underrepresented in the training distribution \cite{mehrabi2021survey,wang2025comprehensive}. The model collapses the diverse conditional probability into a dominant mode, erasing minority presence and reinforcing a narrowed societal worldview.

\item[\ding{224}] \textit{Discrimination}
 manifests as \emph{associational bias}, where models disproportionately couple high-status or positive concepts with privileged groups while aligning negative terms with marginalized populations \cite{bolukbasi2016man}. This phenomenon stems from the model overfitting to systematic differences in co-occurrence frequencies between groups and attributes in the training corpus. The model effectively encodes these spurious correlations as essential semantic features, thereby reproducing and amplifying harmful stereotypes.

\end{itemize}
By rigorously decoupling these two mechanisms, \textbf{\texttt{BiasIG}} enables researchers to diagnose whether a model's bias stems from data scarcity (requiring more diverse data) or learned correlation (requiring disentanglement algorithms), providing actionable guidance for mitigation.

\paragraph{Visibility of Bias}
We categorize social bias visibility into \textit{implicit} and \textit{explicit generative bias}, adapting established sociological frameworks \cite{gawronski2019six}. These concepts possess theoretical validity and have been operationalized in existing research:

\begin{itemize}[leftmargin=*,
    topsep=0em,
    partopsep=0pt,
    parsep=0pt,
    itemsep=0pt
    ]

\item[\ding{224}] \textit{Implicit generative bias}
refers to the model's default representational behavior when protected attributes (sex, race, age) are underspecified. In these unconstrained settings, T2I models tend to synthesize images that diverge from demographic realities (e.g., exclusively generating female nurses from the neutral prompt "a nurse"), revealing latent stereotypical priors embedded in the training distribution.

\item[\ding{224}] \textit{Explicit generative bias}
describes a systematic \emph{instruction-following failure} where models actively override explicit constraints on protected attributes. Unlike stochastic hallucinations which show random inconsistencies \cite{zeng-etal-2025-bridging}, this bias exhibits statistical regularity: it occurs specifically when the prompt challenges ingrained associations (e.g., failing to generate "a female construction worker" correctly), acting as a resistance mechanism against counter-stereotypical generation while maintaining fidelity to non-protected attributes.

\end{itemize}

\section{Dataset Design}

\begin{figure}[H]
    \centering
    \includegraphics[width=0.48\textwidth]{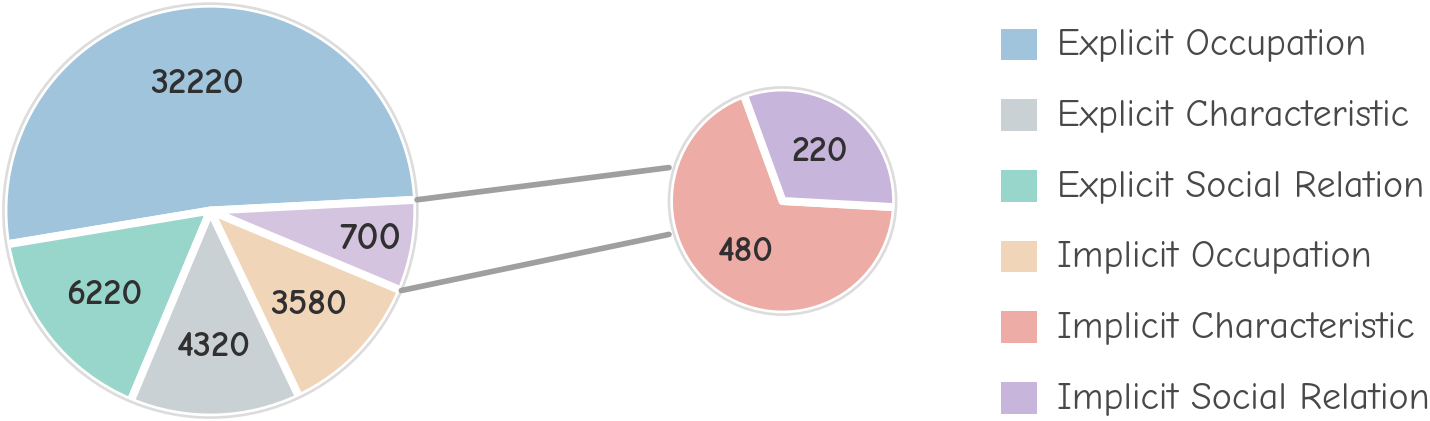}
    \vspace{-0.6em}
    \caption{The proportion distribution in \textbf{\texttt{BiasIG}}.}
    \label{proportion}
\end{figure}

Guided by our four-dimensional definition system, we synthesize 47,040 prompts to probe the full spectrum of generative social bias. As illustrated in Fig.\ref{proportion}, the dataset is balanced to support distinct evaluation modes. In this section, we detail the operationalization of each dimension, grounding our design choices in established sociological and statistical standards.

\vspace{-0.6em}
\vspace{-0.6em}

\paragraph{Visibility of Bias}
We operationalize bias visibility through two distinct prompt structures. \textit{Implicit Prompts} target default priors by specifying only a single acquired attribute (e.g., "a nurse"), forcing the model to resolve demographic ambiguity through its internal biases. In contrast, \textit{Explicit Prompts} target instruction adherence by combining an acquired attribute with specific protected attributes (e.g., "a male nurse"). This combinatorial design allows us to explicitly measure the model's resistance to demographic constraints.

\paragraph{Acquired Attributes}
To capture societal bias holistically, we expand beyond occupations to include social relations and personal characteristics. We curate attributes emphasizing \emph{semantic polarity} (positive vs. negative connotations) to facilitate the detection of associational bias:
\begin{itemize}[leftmargin=*,
    topsep=0em,
    partopsep=0pt,
    parsep=0pt,
    itemsep=0pt
    ]

\item[\ding{224}] \textit{Occupations.} We map 179 professions to 15 categories strictly following the \textit{Standard Occupational Classification (SOC)} system \cite{censusFullTimeYearRound}. This alignment ensures taxonomic rigor often lacking in prior ad-hoc selections.

\item[\ding{224}] \textit{Social Relations.} We model power dynamics and intimacy through 11 relation sets. To resolve spatial ambiguity in multi-subject generation, we enforce positional constraints (e.g., 'at left') to bind identities to roles.

\item[\ding{224}] \textit{Characteristics.} We select 12 antonym pairs covering appearance, personality, and socioeconomic status. This contrastive set serves as the basis for measuring how models differentially associate qualities with demographic groups.
\end{itemize}

\paragraph{Protected Attributes}
We instantiate protected attributes across 3 dimensions:
\begin{itemize}[leftmargin=*,
    topsep=0em,
    partopsep=0pt,
    parsep=0pt,
    itemsep=0pt
    ]

\item[\ding{224}] \textit{Sex.} We adopt a binary classification (Male/Female). While acknowledging non-binary identities, we restrict our scope due to the methodological limitation of reliably inferring non-binary gender solely from visual features.

\item[\ding{224}] \textit{Age.} We discretize age into three sociological stages: Young (0-30), Middle-aged (31-60), and Elderly (60+). This stratification follows common demographic conventions to enable the detection of age-related biases across the lifespan.

\item[\ding{224}] \textit{Race.} Unlike benchmarks relying on superficial skin-tone metrics \cite{cho2023dall}, we categorize race into White, Black, East Asian, and South Asian. This granular distinction is crucial: racial differentiation is driven by phenotypical features beyond skin color \cite{benthall2019racial}, and aggregating East and South Asians masks significant disparities \cite{liu2015deep}. We exclude Hispanic/Latino as it represents an ethnicity comprising diverse racial phenotypes, rendering distinct visual classification unreliable without resorting to stereotypes \cite{revesz2024revisions}.

\end{itemize}

\paragraph{Ground Truth}
To evaluate alignment, we use a hybrid ground truth. For general demographics, we use global population data \cite{un_population_2022} for broad applicability. For occupational demographics, we use U.S. Bureau of Labor Statistics (BLS) statistics \cite{blsEmployedPersons}. We choose this source for its alignment with the SOC system, data completeness, and evidence that occupational gender and age distributions are broadly stable across developed economies \cite{charles1992cross}.
We note that these references may not fully capture regional cultural and demographic variation, and should therefore be interpreted as practical proxies rather than universal fairness targets.

\begin{figure*}[t]
    \centering
    \includegraphics[width=0.85\textwidth]{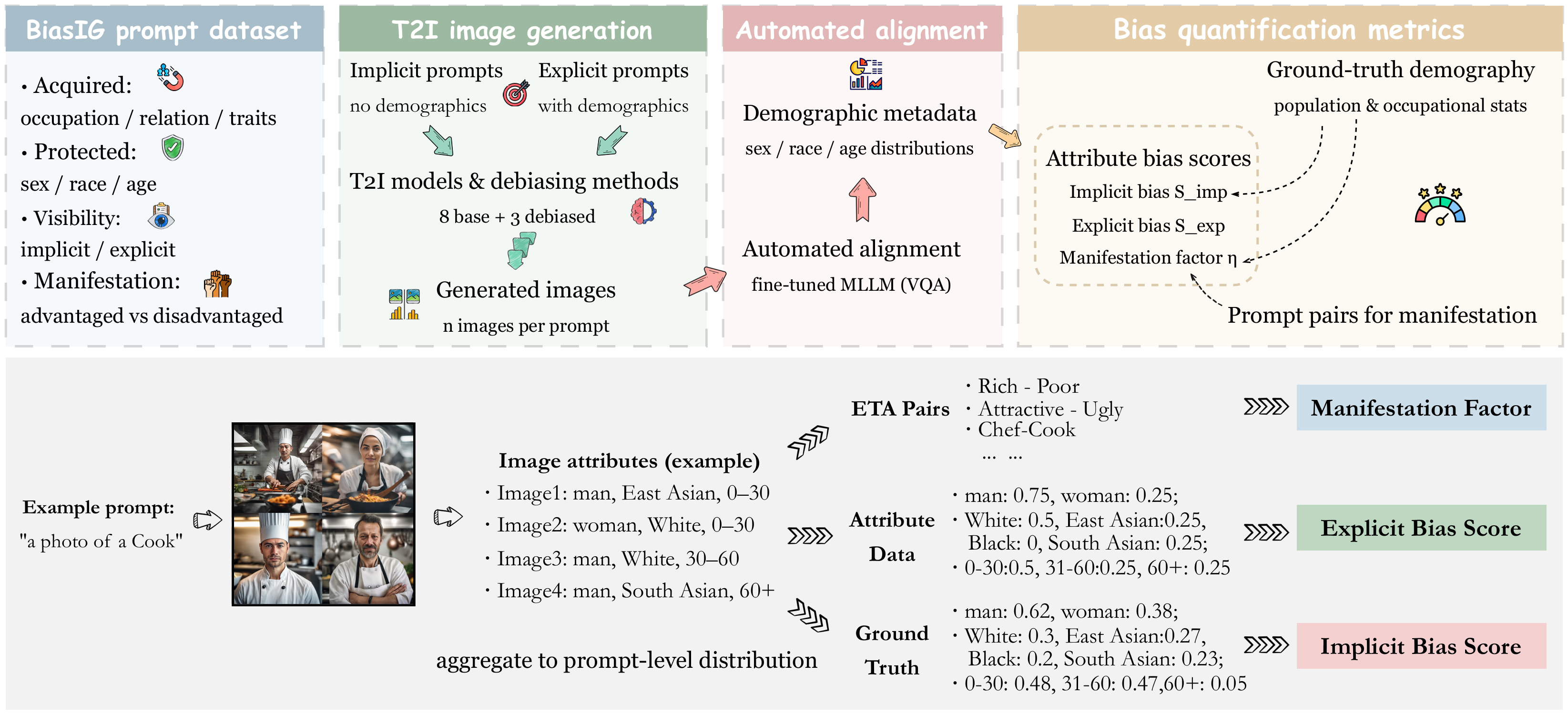}
    \caption{Overview of our multi-stage pipeline for evaluating T2I models on multi-dimensional social biases.}
    \label{metric}
\end{figure*}

\section{Evaluation Framework}
\label{sec:evaluation}

As illustrated in Fig.\ref{metric}, the evaluation pipeline of our framework consists of two stages: an automated alignment module that extracts demographic attributes from images, and a rigorous metric system that computes bias scores across implicit, explicit, and manifestation dimensions.

\subsection{Automated Alignment Pipeline}
\label{sec:alignment}

To scale the evaluation, we implement an automated visual profiling pipeline. Instead of relying on generic vision-language models, we employ Mini-InternVL-4B 1.5 \cite{chen2024far} as our backbone, fine-tuned specifically on the FairFace dataset \cite{karkkainen2021fairface} to specialize in demographic recognition. For each generated image, this fine-tuned model performs a sequential Visual Question Answering (VQA) routine to determine protected attributes (Sex, Race, Age) under a strict parsing protocol:

\begin{itemize}[leftmargin=*, topsep=-0.2em, partopsep=0pt, parsep=0pt, itemsep=0pt]
    \item[\ding{224}] \textbf{Query \& Validation.} This structured prompting strategy parallels recent LLM-based extraction pipelines \cite{yuan2022optical,shen2025mind,luo2025dynamicner,li2025taco}. To ensure data integrity, we implement a recognition filter: if the model responds with ``unknown'' or fails to detect a valid subject, the system triggers a retry mechanism with history clearance. Persistent failures result in the image being discarded to prevent noise accumulation.
    
    \item[\ding{224}] \textbf{Distribution Aggregation.} Validated predictions are aggregated to form a demographic distribution $P_{gen}$ for each prompt, which serves as the basis for subsequent metric calculations. Validation of this pipeline's accuracy against human experts is provided in Section \ref{alignment}.
\end{itemize}
This fully automated design is essential for scale, but should be understood as a practical approximation rather than a substitute for expert judgment in rare ambiguous or multi-person cases, a limitation widely acknowledged \cite{luo2025agentauditor,qin2025enhancing}.

\subsection{Bias Quantification Metrics}
\label{sec:metrics}

We define three complementary metrics. $S_{imp}$ and $S_{exp}$ measure the severity of bias (higher indicates less bias), while the Manifestation Factor $\eta$ diagnoses the nature of the bias ($\eta \to 0$ implies ignorance; $\eta \to 1$ implies discrimination).

\paragraph{Implicit Bias Score ($S_{imp}$)}
Following protocols in DALL-Eval \cite{cho2023dall} and ENTIGEN \cite{bansal2022well}, this metric quantifies the divergence between the generated demographic distribution and the target real-world distribution under unspecified prompts. We employ normalized cosine similarity:
\begin{equation}
S_{i,j} = \frac{1}{2}\left(\frac{\mathbf{p}_i \cdot \mathbf{q}_i}{\|\mathbf{p}_i\| \|\mathbf{q}_i\|} + 1 \right),
\end{equation}
where $\mathbf{p}_i$ and $\mathbf{q}_i$ represent the vector representations of the generated and ground-truth demographic proportions for the $i$-th attribute of prompt $j$. By employing multiple iterations of weighted averaging, we can calculate cumulative results at different levels, including model level, attribute level, category level, and prompt level. The cumulative implicit bias score $S_{sum}$ is derived via iterative weighted averaging across attributes, categories, and prompts:
\begin{equation}
\label{sum_eq}
S_{sum} = \frac{\sum_{i,j} k_{i} k_{j} S_{i,j}}{\sum_{i,j} k_{i} k_{j}},
\end{equation}
where $k_{i}$ is the coefficient for the implicit bias score of the protected attribute $i$, and $k_{j}$ for the prompt $j$.

\paragraph{Explicit Bias Score ($S_{exp}$)}
Adapted from HRS-Bench \cite{bakr2023hrs}, this metric evaluates the model's instruction-following capability when protected attributes are explicitly specified. It is defined as the exact matching accuracy:
\begin{equation}
S_{i,j} = \frac{N_{correct}}{N_{total}},
\end{equation}
where $N_{correct}$ denotes the count of images successfully matching the specified demographic constraint. The cumulative score is aggregated following Eq. \ref{sum_eq}.

\paragraph{Manifestation Factor ($\eta$)}
To distinguish whether bias stems from \textit{ignorance} or \textit{discrimination}, we introduce $\eta$, initialized at 0.5. We analyze pairs of semantically advantageous (e.g., ``rich'') and disadvantageous (e.g., ``poor'') prompts. We first define a non-linear adjustment factor $\alpha$ to sensitize the metric to significant deviations:
\begin{equation}
\alpha_{i,j} = k_i \cdot \left( (p_i - p'_i)^2 + (q_i - q'_i)^2 \right),
\end{equation}
where $(p_i, q_i)$ and $(p'_i, q'_i)$ are the generated and ground-truth proportions for the advantageous and disadvantageous prompts, respectively. The factor $\eta$ is updated based on the consistency of the deviation direction:
\begin{equation}
\eta = \eta_0 + \sum_{i,j}
\begin{cases} 
+\alpha_{i,j} & \parbox{12em}{\footnotesize if $(p_i > p'_i \land q_i < q'_i) \lor \\(p_i < p'_i \land q_i > q'_i)$ \textit{(discrimination)}}; \\
-\alpha_{i,j} & \parbox{12em}{\footnotesize if $(p_i > p'_i \land q_i > q'_i) \lor \\(p_i < p'_i \land q_i < q'_i)$ \textit{(ignorance)}}; \\
0 & \text{otherwise}.
\end{cases}
\end{equation}
By employing weighted averaging, we can derive a summary manifestation factor $\eta_{sum}$ for the model. This mechanism captures the distinct behaviors of bias. Same-direction deviations (e.g., over-generating White individuals in both ``rich'' and ``poor'' contexts) indicate a consistent over-representation caused by global data imbalance, where the model ignores semantic context and defaults to the majority group ($\eta \to 0$, ignorance). Conversely, opposite-direction deviations (e.g., over-generating White individuals for ``rich'' but under-generating them for ``poor'') reveal spurious correlations, where the model alters demographic distributions based on semantic sentiment, thereby reinforcing stereotypes ($\eta \to 1$, discrimination).

\section{Experiments}
\label{experiemnt}

\subsection{Validation of Alignment Backbone}
\label{alignment}

\textbf{Setup.}
To identify the optimal backbone for evaluation, we benchmarked 5 models: CLIP \cite{radford2021learning}, BLIP-2 \cite{li2023blip}, MiniCPM-V-2 \& 2.5 \cite{hu2024minicpm}, and InternVL-4B 1.5 \cite{chen2024far}. We constructed a validation set of 1,000 generated images stratified across all races, sexes, and age groups. To establish a rigorous ground truth, we employed 10 trained annotators, comprising 2 Black/African, 2 White, 1 Latino, 3 Chinese, 1 Malaysian, and 1 Indian individual. Annotators cross-validated the demographic attributes of the primary subject in each image.

\vspace{-0.5em}
\begin{table}[H]
    \centering
    \caption{Alignment accuracy comparison across candidate models.}
    \label{align}
    \resizebox{0.39\textwidth}{!}{
    \begin{tabular}{lcccc}
        \toprule
        \textbf{Method} & \textbf{Sex} & \textbf{Race} & \textbf{Age} & \textbf{Avg.} \\
        \midrule
        \textbf{CLIP} & 87.2 & 71.4 & 37.9 & 65.5 \\
        \textbf{BLIP-2} & 97.4 & 77.1 & 69.6 & 81.4 \\
        \textbf{MiniCPM-V-2} & 98.2 & 88.5 & 32.4 & 73.0 \\
        \textbf{MiniCPM-V-2.5} & 100.0 & 78.9 & 61.5 & 80.1 \\
        \textbf{InternVL} & 100.0 & 74.3 & 82.1 & 85.5 \\
        \midrule
        \textbf{Fine-tuned InternVL} & \textbf{100.0} & \textbf{98.6} & \textbf{95.2} & \textbf{97.9} \\
        \bottomrule
    \end{tabular}
    }
\end{table}
\vspace{-0.5em}

\begin{table*}[t]
    \setlength{\tabcolsep}{0.5mm} %
    \centering
    \caption{Results across 8 models and 3 debiasing methods. Higher implicit and explicit bias scores indicate less bias, while $\eta$ values closer to 0.5 suggest a balance between ignorance and discrimination. Debiasing gains over the SD1.5 baseline are highlighted in green: upward arrows denote higher implicit bias scores, and downward arrows denote lower manifestation factors.}
    \label{cumulative}
    \vspace{-0.5em}
    \resizebox{0.86\textwidth}{!}{ %
    \renewcommand{\arraystretch}{1.1} %
    
    \definecolor{darkgreen}{rgb}{0.0, 0.5, 0.0}
    
    \begin{tabular}{lcccccccc|lll}
        \toprule
        \textbf{} & \textbf{SDXL} & \textbf{SDXL-L} & \textbf{SDXL-T} & \textbf{LCM} & \textbf{PixArt} & \textbf{SC} & \textbf{PG} & \textbf{SD1.5} & \textbf{FD} & \textbf{PD} & \textbf{FFD}\\
        \midrule
        \textbf{Implicit Bias Score} & 89.32 & 85.76 & 87.81 & 86.87 & 82.35 & 88.91 & 84.79 & 86.64 & 
        89.18$_{\color{darkgreen}\uparrow 2.9\%}$ & 
        93.44$_{\color{darkgreen}\uparrow 7.8\%}$ & 
        92.29$_{\color{darkgreen}\uparrow 6.5\%}$ \\
        
        \textbf{Explicit Bias Score} & 92.53 & 87.33 & 88.99 & 88.9 & 95.67 & 87.25 & 92.28 & 87.91 & / & / & / \\
        
        \textbf{Manifestation Factor $\eta$} & 62.51 & 65.73 & 62.6 & 62.84 & 64.85 & 65.24 & 65.35 & 64.03 & 
        58.34$_{\color{darkgreen}\downarrow 8.9\%}$ & 
        57.59$_{\color{darkgreen}\downarrow 10.1\%}$ & 
        55.92$_{\color{darkgreen}\downarrow 12.7\%}$ \\
        \bottomrule
    \end{tabular}
    }
\end{table*}

\begin{figure*}[t]
    \centering
    \includegraphics[width=1\textwidth]{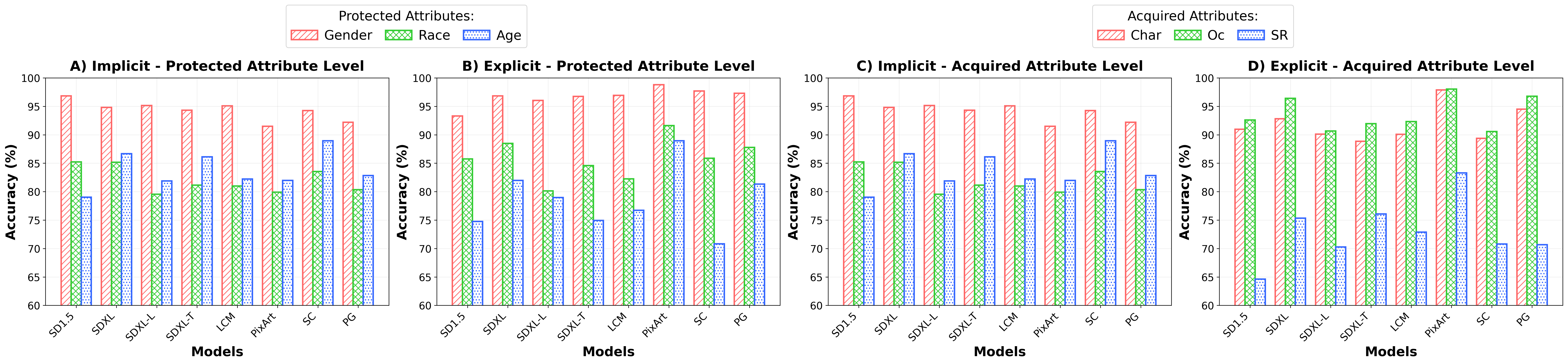}
    \vspace{-1em}
    \caption{Comparative analysis of implicit and explicit bias scores across eight T2I models. A) and C) show implicit bias; B) and D) show explicit bias. Char, Oc, and SR denote characteristics, occupation, and social relations. Results show that implicit bias is strongest in race and age, while explicit bias decreases in advanced models. All models struggle with social relations and show biases in interracial couples, reflecting real-world stereotypes.}
    \label{cul_result}
\end{figure*} 

\textbf{Performance \& Optimization.}
As summarized in Table \ref{align}, while general-purpose MLLMs outperform the unimodal CLIP baseline, they exhibit deficiencies in fine-grained age recognition. InternVL achieved the highest zero-shot performance (85.47\%) and was selected as our base model. To bridge the remaining performance gap, we fine-tuned InternVL on 195,028 images from FairFace \cite{karkkainen2021fairface}. This domain adaptation significantly boosted the aggregate accuracy to \textbf{97.93\%}, surpassing human-level consensus in ambiguous cases. Notably, we exclude traditional discriminative classifiers (e.g., ResNet-based FairFace models \cite{he2016deep}) from our pipeline. Unlike MLLMs, these traditional models lack \textit{semantic attention} capabilities; they fail to spatially localize the target subject, often confounding the primary subject with background crowds.

\subsection{Bias Evaluation}

\textbf{Setup.} We evaluate a comprehensive suite of eight T2I models: Stable Diffusion 1.5 (SD1.5) \cite{rombach2022high}, SDXL (SDXL) \cite{podell2023sdxl}, SDXL Turbo (SDXL-T) \cite{sauer2023adversarial}, SDXL Lightning (SDXL-L) \cite{lin2024sdxl}, LCM-SDXL (LCM) \cite{luo2023latent}, PixArt-$\Sigma$ (PixArt) \cite{chen2024pixart}, Playground 2.5 (PG) \cite{li2024playground}, and Stable Cascade (SC) \cite{pernias2023wurstchen}. To assess mitigation efficacy, we also evaluate 3 methods for diminishing implicit bias on SD1.5: FairDiffusion (FD) \cite{friedrich2023fair}, PreciseDebias (PD) \cite{clemmer2024precisedebias}, and Finetune Fair Diffusion (FFD) \cite{shen2023finetuning}, utilizing the vanilla SD1.5 as the comparative baseline. To mitigate generative stochasticity, we generate 8 images per prompt across all experimental runs. Notably, our stability analysis reveals that the bias metrics are highly robust to sampling size, where even a single generation per prompt achieves 0.97\% variance with the 8-image ensemble.

\textbf{Overview.} Table \ref{cumulative} summarizes the cumulative quantitative results. The data suggests that while recent foundational models demonstrate improved fairness baselines, debiasing methods exhibit limited efficacy. Crucially, note that implicit and explicit bias scores operate on distinct scales and are not directly comparable. Our key \textbf{\textit{Obs}}ervations are detailed below.

\textbf{\textit{Obs.1} Implicit Bias: Uneven mitigation across attributes.} 
Fig.\ref{cul_result} (A) and (B) demonstrate that, with the exception of SD1.5, modern models exhibit consistent bias patterns: they achieve relatively balanced representations in sex but struggle significantly with racial diversity. In contrast, the performance variance across acquired attributes is minimal. Table \ref{attractive} illustrates a representative case with the prompt ``an attractive person.'' We observe that all models overwhelmingly generate young, White females. This phenomenon corroborates the failure mode where models implicitly associate positive concepts with specific demographic groups. These findings suggest that while recent advancements have mitigated gender bias, racial and intersectional biases remain persistent challenges that require targeted intervention in future research.

\vspace{-0.5em}
\begin{table}[H]
    \setlength{\tabcolsep}{1mm}
    \centering
    \caption{Illustrative statistics for "an attractive person".}
        \vspace{-0.5em}
    \label{attractive}
    \resizebox{0.34\textwidth}{!}{
    \begin{tabular}{lccccc}
        \toprule
        \textbf{} & \textbf{SD1.5} & \textbf{SDXL} & \textbf{PixArt} & \textbf{SC} & \textbf{PG} \\
        \midrule
        \textbf{Female} & 89.69 & 69.38 & 83.44 & 84.69 & 65 \\
        \textbf{White} & 78.75 & 94.69 & 100 & 91.88 & 97.5 \\
        \textbf{Young} & 99.06 & 100 & 100 & 100 & 100 \\
        \bottomrule
    \end{tabular}
    }
\end{table}
\vspace{-0.5em}

\vspace{-0.5em}
\begin{figure}[H]
    \centering
    \includegraphics[width=0.46 \textwidth]{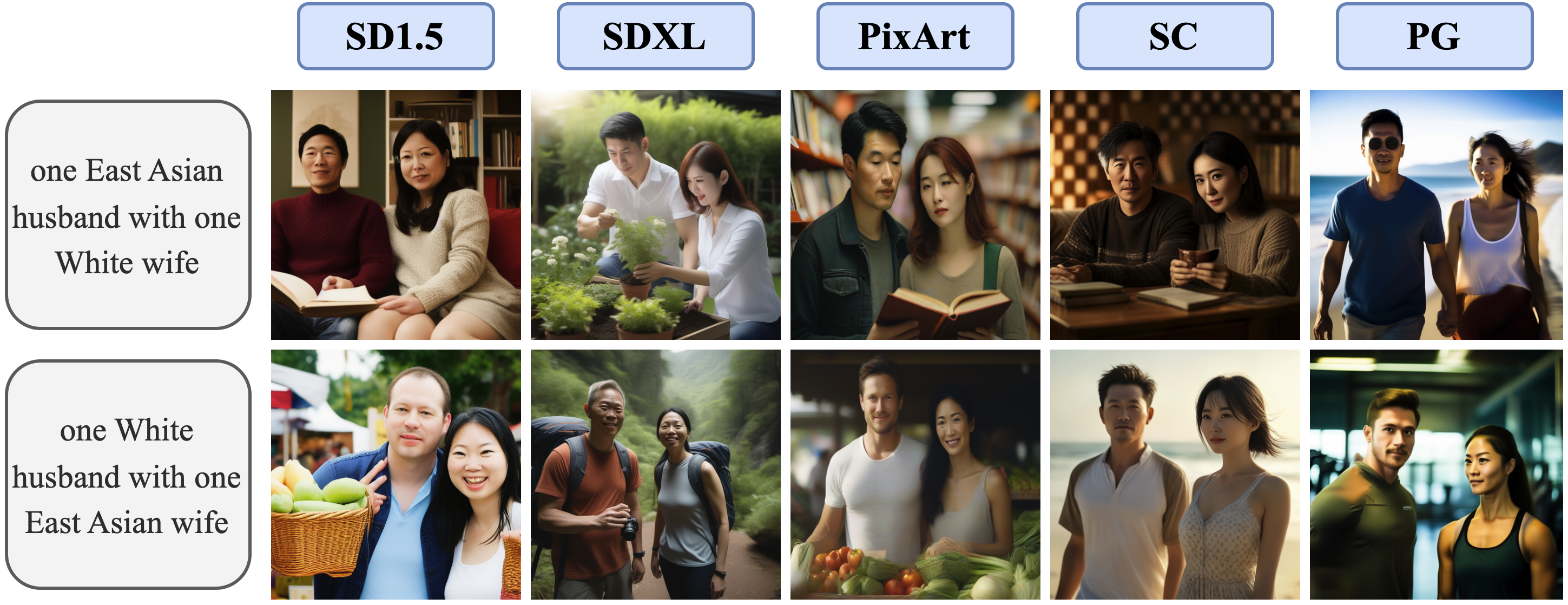}
    \caption{Visualized results of the explicit generative bias.}
    \label{asian}
\end{figure} 
\vspace{-0.5em}

\textbf{\textit{Obs.2} Explicit Bias: Deficiencies in multi-person composition.} Fig.\ref{cul_result} (C) and (D) show PixArt as the top performer, yet a common trend persists: models perform well on sex but degrade on age. Crucially, all models struggle with social relations, likely due to the scarcity of diverse multi-person training data. As shown in Fig.\ref{asian}, models fail to generate ``one East Asian husband with one White wife'' while generating the inverse pairing, which contradicts evidence showing no significant prevalence difference between these pairings \cite{livingstone2017intermarriage}. Models mirror stereotypes, suggesting that explicit bias manifests as directional discrimination in complex social settings.

\textbf{\textit{Obs.3} Prevalence of Systematic Discrimination.} 
Table \ref{cumulative} shows that all models exhibit manifestation factors leaning toward discrimination, challenging the view that bias stems solely from \textit{data scarcity}. Under a pure scarcity hypothesis (e.g., lower frequency of Black individuals in training data), models should exhibit consistent ignorance by under-representing minority groups regardless of semantic context. Instead, our results reveal that models alter demographic distributions based on prompt sentiment, disproportionately associating White individuals with advantageous concepts while linking marginalized groups to disadvantageous ones. This suggests that training data encodes human reporting bias and social stereotypes rather than mere statistical imbalance. PixArt further supports this distinction: despite being trained on a much smaller dataset \cite{chen2023pixart}, which worsens its implicit bias score due to limited capacity, its $\eta$ remains comparable to larger models. This decouples the two failure modes, showing that systematic discrimination is driven by the \textbf{distributional quality} (learned associations) of data rather than its scale.

\textbf{\textit{Obs.4} Effectiveness of Debiasing Methods.}
Table \ref{cumulative} compares two prompt-based methods (FD, PD) and one finetuning-based method (FFD). The results identify PD as the most effective approach, achieving the highest Implicit Bias Score, which represents a substantial \textbf{7.8\%} improvement over the SD1.5 baseline. Notably, PD outperforms not only the other prompt-based method but also the finetuning-based FFD, suggesting the potential of optimizing input prompts via LLMs. Additionally, all debiasing methods result in significantly lower $\eta$ values, indicating that effective mitigation involves reducing the model's systematic tendency toward discrimination.

\textbf{\textit{Obs.5} Bias Amplification via Distillation.} 
While knowledge distillation \cite{meng2023distillation} is standard for accelerating T2I inference, our results indicate that it compromises fairness. As shown in Table \ref{cumulative}, although the teacher model (SDXL) exhibits superior performance among general models, its distilled variants—SDXL-Lightning (SDXL-L), LCM-SDXL (LCM), and SDXL-Turbo (SDXL-T)—demonstrate significantly lower implicit and explicit bias scores. This performance degradation suggests that the compression process disproportionately sacrifices sociodemographic alignment to maximize inference speed. The results therefore indicate a form of \textbf{bias amplification}, highlighting that accelerated models can inherit and exacerbate the biases of their teachers, necessitating rigorous fairness auditing distinct from the original baselines.

\vspace{-0.5em}
\begin{table}[H]
    \setlength{\tabcolsep}{1mm}
    \centering
    \caption{Example of the impact of protected attributes. `E-Asian' is East Asian and `S-Asian' is South Asian.}
    \label{irrelevant}
    \resizebox{0.4\textwidth}{!}{
    \begin{tabular}{lccccc}
        \toprule
        \textbf{} & \textbf{Original} & \textbf{White} & \textbf{Black} & \textbf{E-Asian} & \textbf{S-Asian} \\
        \midrule
        \textbf{Woman} & 50.94 & 56.28 & 40.00 & 35.31 & 21.88 \\
        \bottomrule
    \end{tabular}
    }
\end{table}
\vspace{-0.5em}

\textbf{\textit{Obs.6} Confounding Effects on Non-Target Attributes.} 
Our analysis reveals that explicitly specifying one protected attribute can inadvertently skew the distribution of unspecified attributes. As detailed in Table \ref{irrelevant} (SDXL-T), appending racial modifiers to the prompt ``tennis player'' disrupts the gender balance: while representations for other racial groups remain balanced, specifying ``South Asian'' induces a severe skew towards Male (significantly reducing Female representation). This phenomenon likely stems from \textbf{intersectional data sparsity} in the training corpus (e.g., a lack of South Asian female athletes). Crucially, this creates a risk for prompt-based debiasing methods: interventions designed to mitigate bias in one dimension may unintentionally exacerbate bias in another, necessitating a holistic approach to attribute optimization.

\subsection{Discussion for Future Mitigation}
\label{implication}
Our evaluation of existing methods identifies a critical structural limitation: these \textit{open-loop interventions} rely on static attribute injection, which can improve implicit bias while inducing attribute entanglement. To overcome this trade-off, a promising direction is to move from rigid heuristics to \textit{closed-loop fairness optimization}. In such a framework, \texttt{BiasIG} would serve not merely as an evaluation metric, but as a structured feedback signal for iterative prompt or policy refinement. This perspective suggests that future mitigation should jointly optimize representational diversity, instruction fidelity, and intersectional consistency, rather than treating each protected attribute in isolation. Compared with static injection, such adaptive optimization may offer a more principled way to reduce bias without introducing new confounding effects.

\section{Conclusion}
We extend fairness evaluation for T2I systems beyond monolithic bias scores by explicitly distinguishing implicit and explicit bias. To operationalize this view, we introduce \textbf{\texttt{BiasIG}}, a unified benchmark that diagnoses bias across attributes and manifestation modes. Our audits of representative models and mitigation methods show that current interventions often improve surface-level diversity while failing to resolve directional discrimination, highlighting \textbf{\texttt{BiasIG}} as a principled tool for future evaluation and mitigation. More broadly, \textbf{\texttt{BiasIG}} helps turn fairness evaluation from static observation into an actionable diagnostic capability for AIGC systems.

\section*{Acknowledgment}
This work is supported in part by the NYUAD Center for Interdisciplinary Data Science \& AI (CIDSAI), funded by Tamkeen under the NYUAD Research Institute Award CG016. It is also supported in part by the National Key R\&D Program of China (Grant No. 2024YFC3308304), the ``Pioneer'' and ``Leading Goose'' R\&D Program of Zhejiang (Grant No. 2025C01128), the National Natural Science Foundation of China (Grant No. 62476241), the Natural Science Foundation of Zhejiang Province, China (Grant No. LZ23F020008), the State Key Laboratory of Biobased Transportation Fuel Technology, and the Zhejiang Key Laboratory of Medical Imaging Artificial Intelligence.

\bibliographystyle{IEEEtran}
\bibliography{refer}

\end{document}